\documentclass[aps,prl,twocolumn,showpacs,superscriptaddress,nofootinbib]{revtex4-1}
\usepackage[margin=1in]{geometry} 

\usepackage{times}
\usepackage[small,compact]{titlesec}

\usepackage{lineno}  
\usepackage{graphicx}  
\usepackage{xspace} 
\usepackage{color}
\usepackage{colortbl}
\usepackage{amsmath} 
\usepackage{ifthen} 

\usepackage{multirow}
\usepackage{booktabs}
\usepackage{longtable}
\usepackage{amsmath}
\usepackage{verbatim}
\usepackage{mciteplus}

\usepackage{wrapfig}

\usepackage{hyperref}
\def\name{Jibo He}

\hypersetup{
  pdfauthor = {\name},
  pdfkeywords = {RICH, calibration, alignment},
  pdftitle = {Real-time calibration and alignment of the LHCb RICH detectors},
  pdfsubject = {Summary},
  pdfpagemode = UseNone
}

\newboolean{pdflatex}
\setboolean{pdflatex}{true} 
\newboolean{articletitles}
\setboolean{articletitles}{true} 

\usepackage{amssymb}
\usepackage{amsfonts}
\usepackage{upgreek} 

\usepackage{hyperref}    
\usepackage[all]{hypcap} 

\usepackage{epstopdf}

\begin{document}
\graphicspath{{figs/}}


\title{Real-time calibration and alignment of the LHCb RICH detectors}


\author{Jibo~HE (on behalf of the LHCb RICH collaboration)}
\affiliation{University of Chinese Academy of Sciences (UCAS),
  Beijing, China}

\begin{abstract}
In 2015, the LHCb experiment established 
a new and unique software trigger strategy 
with the purpose of increasing the purity of the
signal events by applying the same algorithms online and offline. 
To achieve this, 
real-time calibration and alignment of 
all LHCb sub-systems is needed to provide
vertexing, tracking, and particle identification of the best possible
quality. 
The calibration of the refractive index of the
RICH radiators, the calibration of the Hybrid Photon Detector
image, and the alignment of the RICH mirror system, are reported in
this contribution. The stability of the RICH performance
and the particle identification performance are also
discussed. 
\end{abstract}

\maketitle

\section{Introduction}
The LHCb experiment~\cite{Alves:2008zz} is one of the four large
particle physics experiments at the Large Hadron Collider (LHC),
and is designed to search for physics beyond the Standard Model by
precision study of the beauty and charm hadrons.
The Ring Imaging CHerenkov (RICH) detectors of the LHCb
experiment employ the ${\rm C_4 F_{10}}$ and ${\rm C F_{4}}$ radiators
to 
provide particle identification (PID) of charged particles in the
momentum range of 2-100 GeV. This is essential for the LHCb core
physics program. 
The centre-of-mass energy of the LHC has been
increased from 8 TeV to 13 TeV in 2015, and the LHCb experiment will
be upgraded~\cite{LHCb-TDR-012, LHCb-TDR-014}, starting in 2019, 
and run at a 4-times higher luminosity than in LHC Runs I and II.
A new and unique software trigger strategy has been established
at the LHCb experiment with the purpose of increasing the purity of
the signal events by applying the same algorithms online and offline. 
This requires ultimate quality of vertexing, tracking, and PID. 
Therefore, real-time calibration and alignment of all LHCb sub-systems
is required online.  
The calibration of the refractive index of
the RICH radiators, the calibration of the Hybrid Photon Detector
(HPD) image, and the alignment of the RICH mirror system are reported
here. 

\begin{figure}[bt]
  \centering
  \resizebox{0.235\textwidth}{!}{%
    \includegraphics{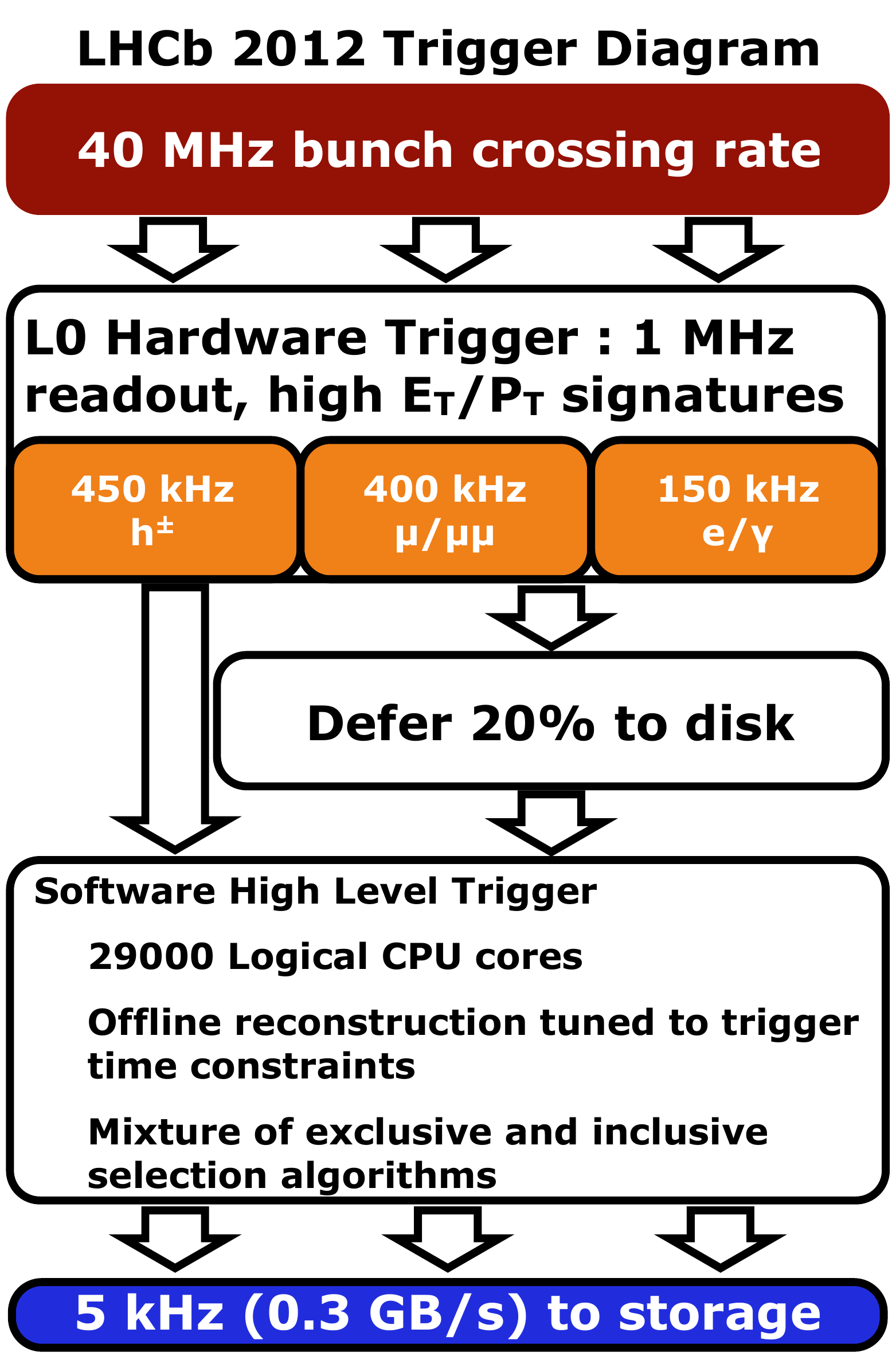}}
  \resizebox{0.235\textwidth}{!}{%
    \includegraphics{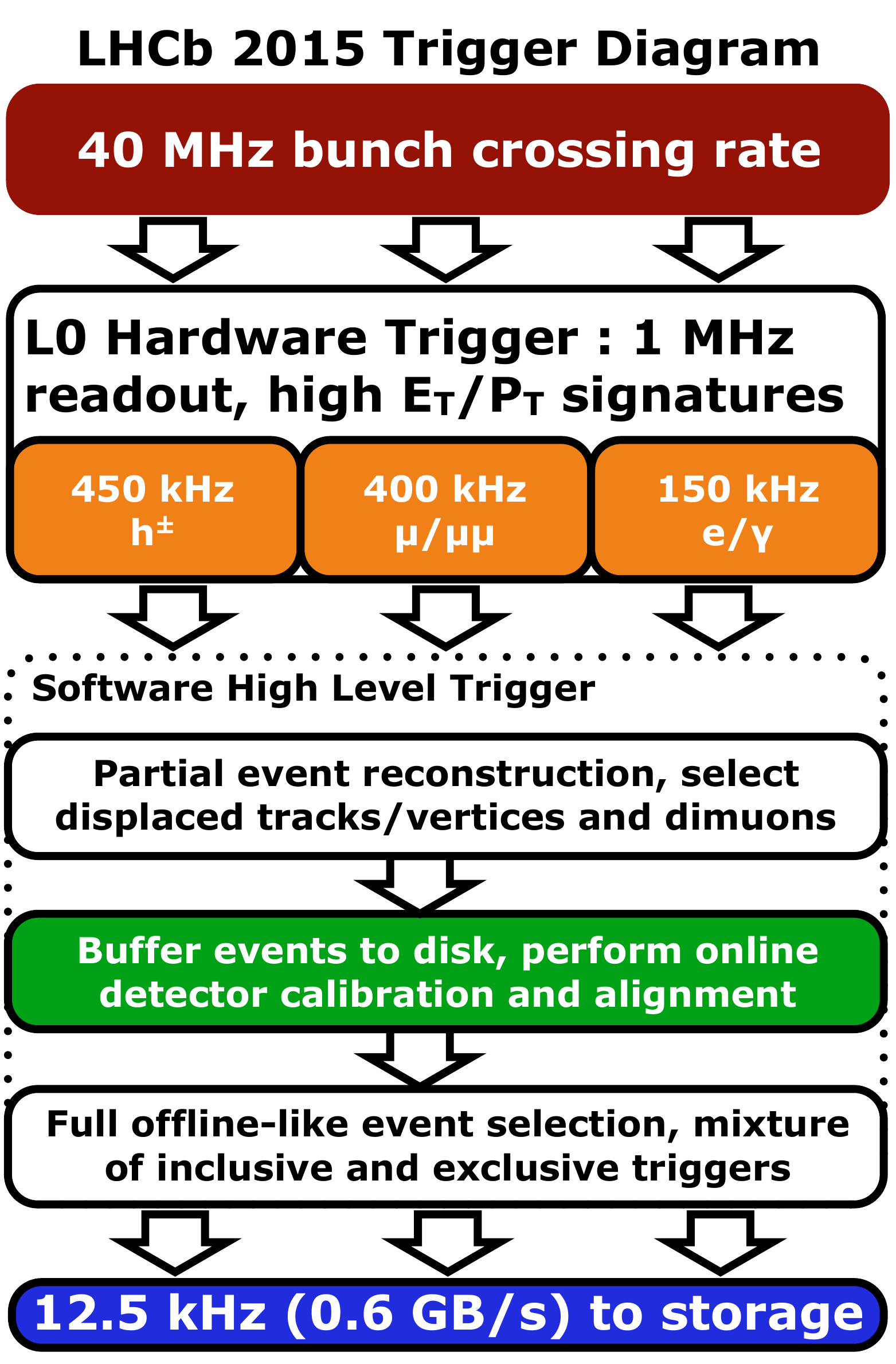}}
  \caption{Schematic diagram of the LHCb trigger data-flow in Run-I
    data taking (left) compared to the data-flow in Run-II (right).}
  \label{fig:TrgDataFlow}
{\vskip-0.2cm}{\vskip-0.2cm}\end{figure}

The LHCb trigger strategy in Run-II (2015-2018) has been changed 
with respect to that in Run-I (2010-2013), as shown in Fig.~\ref{fig:TrgDataFlow}. 
The online event reconstruction in Run-I was simpler and faster than
that used offline, and did not have the latest detector
calibration and alignment constants applied. 
In the Run-II data-taking, the events selected by the first stage of 
the software trigger are buffered on local disks, then an automatic
detector calibration and alignment is performed, and the resulting calibration 
and alignment constants are used in the final stage of the software 
trigger. As the full offline event reconstruction is run 
in the final stage of the software trigger, its output
can be used for physics analysis directly without further offline
processing. 

\section{RICH optical layout and reconstruction}
\begin{figure}
  \centering
  \resizebox{0.3\textwidth}{!}{%
    \includegraphics{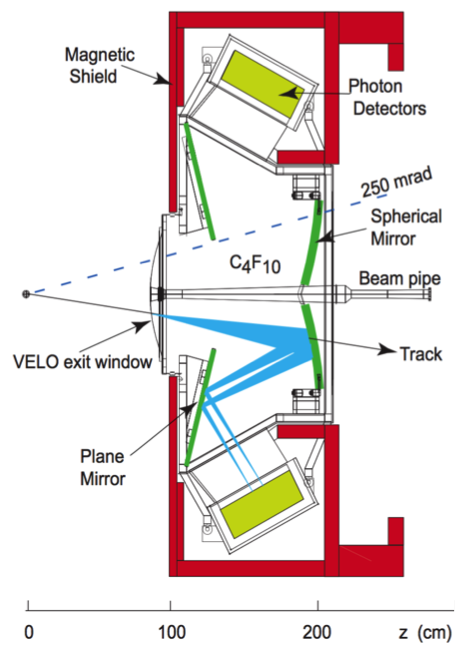}}
  \caption{Side view of the LHCb RICH detector upstream of the magnet.}
  \label{fig:Rich1Sch}
{\vskip-0.4cm}\end{figure}

The LHCb RICH system has two detectors, one upstream and one
downstream of the magnet, covering the 
low momentum range $\sim2-60$~GeV$/c$ and the 
high momentum range $\sim15-100$~GeV$/c$ respectively~\cite{Alves:2008zz}. 
The side view of the RICH detector upstream of the magnet is shown in
Fig.~\ref{fig:Rich1Sch}.
The Cherenkov photons emitted by charged particle tracks in the
radiator are reflected and focused by the combination of spherical and
secondary mirrors out of the LHCb geometrical acceptance and then are detected
by the Hybrid Photon Detectors (HPD). 

To reconstruct the photon candidate and the Cherenkov angle, 
one takes the spatial position of the HPD pixel hits as the detection
point, and the middle point of the
associated track in the radiator as the emission point, then solves a
quartic equation that fully describes the reflections of the photon, 
given the RICH geometry~\cite{CkvAngRec}.  
The ``global likelihood algorithm''~\cite{GlobalPID} is used to
determine the PID, where information from all tracks, 
all radiators and all pixel hits within a given event are considered
simultaneously, and the likelihood is maximized by comparing the
observed Cherenkov angles with those expected under different PID
hypotheses. 

To achieve the best PID performance, one needs to 
align the RICH mirrors, detector planes and the tracking system, 
and calibrate the refractive index of radiators and the HPD image 
with good precision. These factors are all time-dependent, necessitating 
real-time calibration and alignment of
the LHCb RICH detectors, and the tracking system.

\section{Calibration and alignment}

\subsection{Calibration of the refractive index of the RICH radiators}
\begin{figure}
  \centering
  \resizebox{0.4\textwidth}{!}{%
    \includegraphics{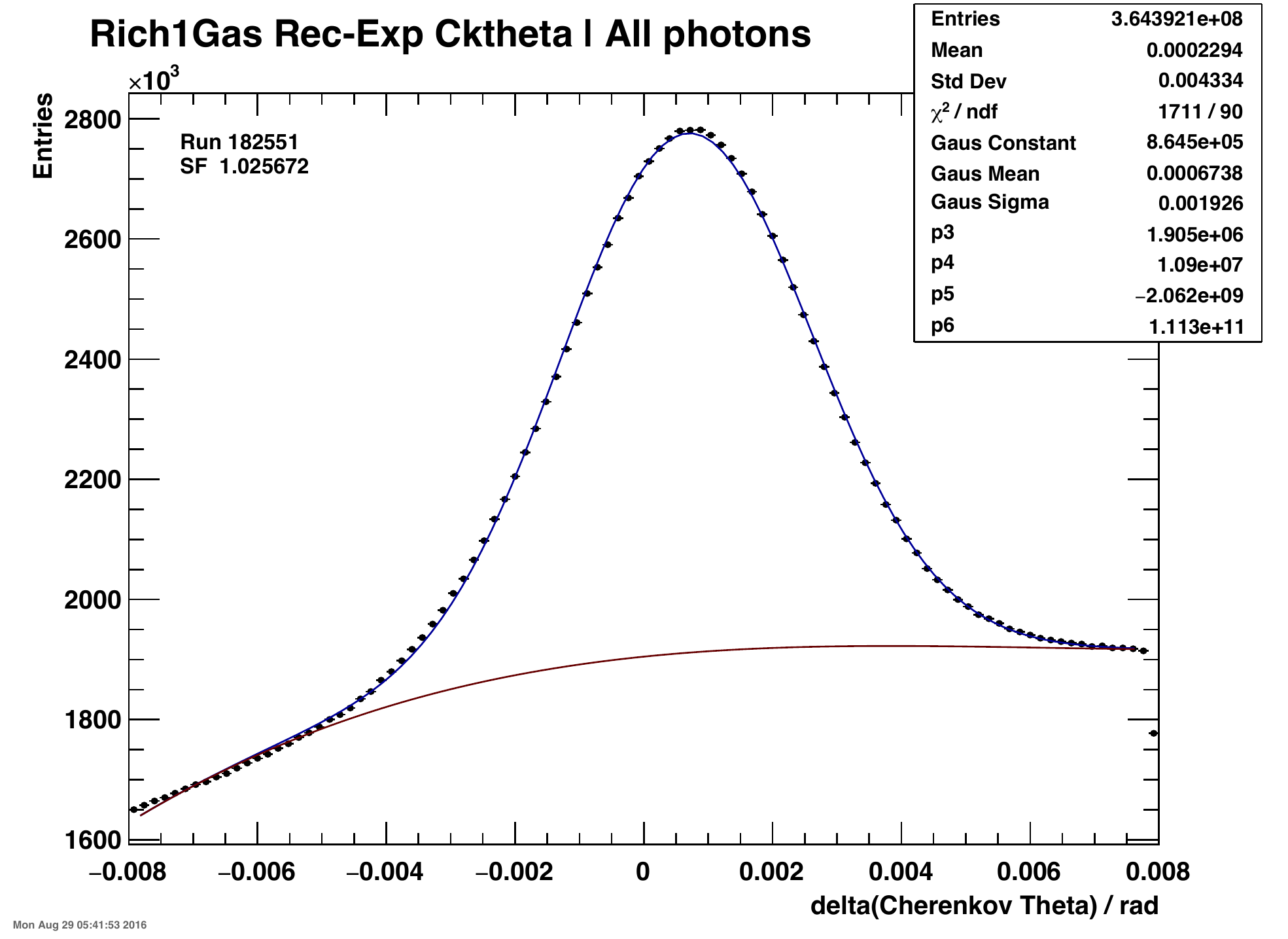}}
  \caption{Difference between the reconstructed and expected Cherenkov
    angle before the calibration.}
  \label{fig:RichCal}
{\vskip-0.2cm}\end{figure}

The refractive index of the gas radiators depends on the ambient
temperature and pressure, and the exact composition of the gas
mixture; so it can change in time. 
These quantities are monitored by hardware 
to compute an expected refractive index, 
but this does not have a precision 
that is high enough for the physics analysis, 
therefore it needs to be further corrected.
As shown in Fig.~\ref{fig:RichCal},  
the distribution of the difference between the
reconstructed and expected Cherenkov angle is fitted
to obtain the shift, which is then converted 
to a scale factor of the expected refractive index
according to studies based on simulation. 

About 50~Hz of events are sent to multiple online reconstruction tasks,
which run in parallel, and the resulting histograms are
merged at the end of each run. Then a dedicated task is used 
to fit the histograms merged run-by-run and produce calibration
constants to be used by the RICH reconstruction in the 
final stage of the software trigger. The maximum run
length is one hour.

\subsection{Calibration of the HPD images}

\begin{figure}
  \centering
\includegraphics[width=7cm]{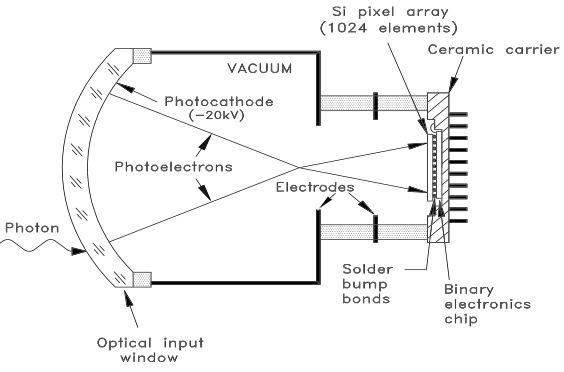}
\caption{Schematic drawing of the Hybrid Photon Detector (HPD).}
\label{fig:HPD}
{\vskip-0.2cm}\end{figure}

\begin{figure}
  \resizebox{0.5\textwidth}{!}{%
    \includegraphics{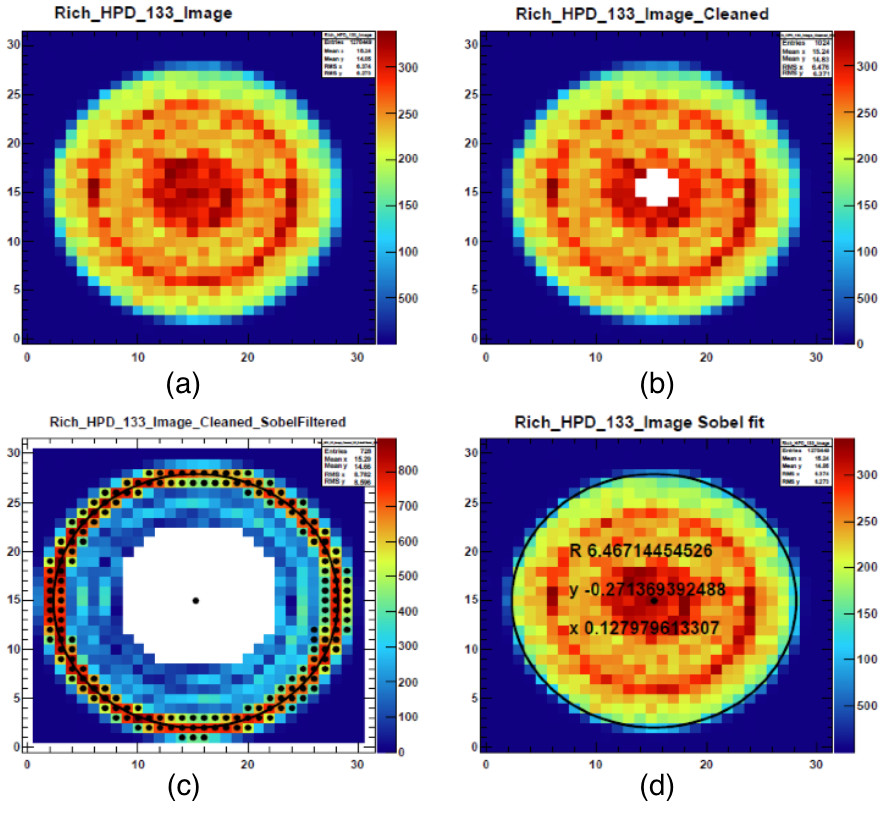}}
  \caption{Calibration process of the HPD image. 
    (a)~HPD image for a typical run. (b)~The centre of the HPD image is
    cleaned to eliminate ion feedback. (c)~Edge of the HPD image detected by the
    Sobel filter. (d)~Radius and centre of the HPD image returned by the fit. 
  }
  \label{fig:RichHPDCal}
\vskip0.5cm
\resizebox{0.45\textwidth}{!}{%
\includegraphics{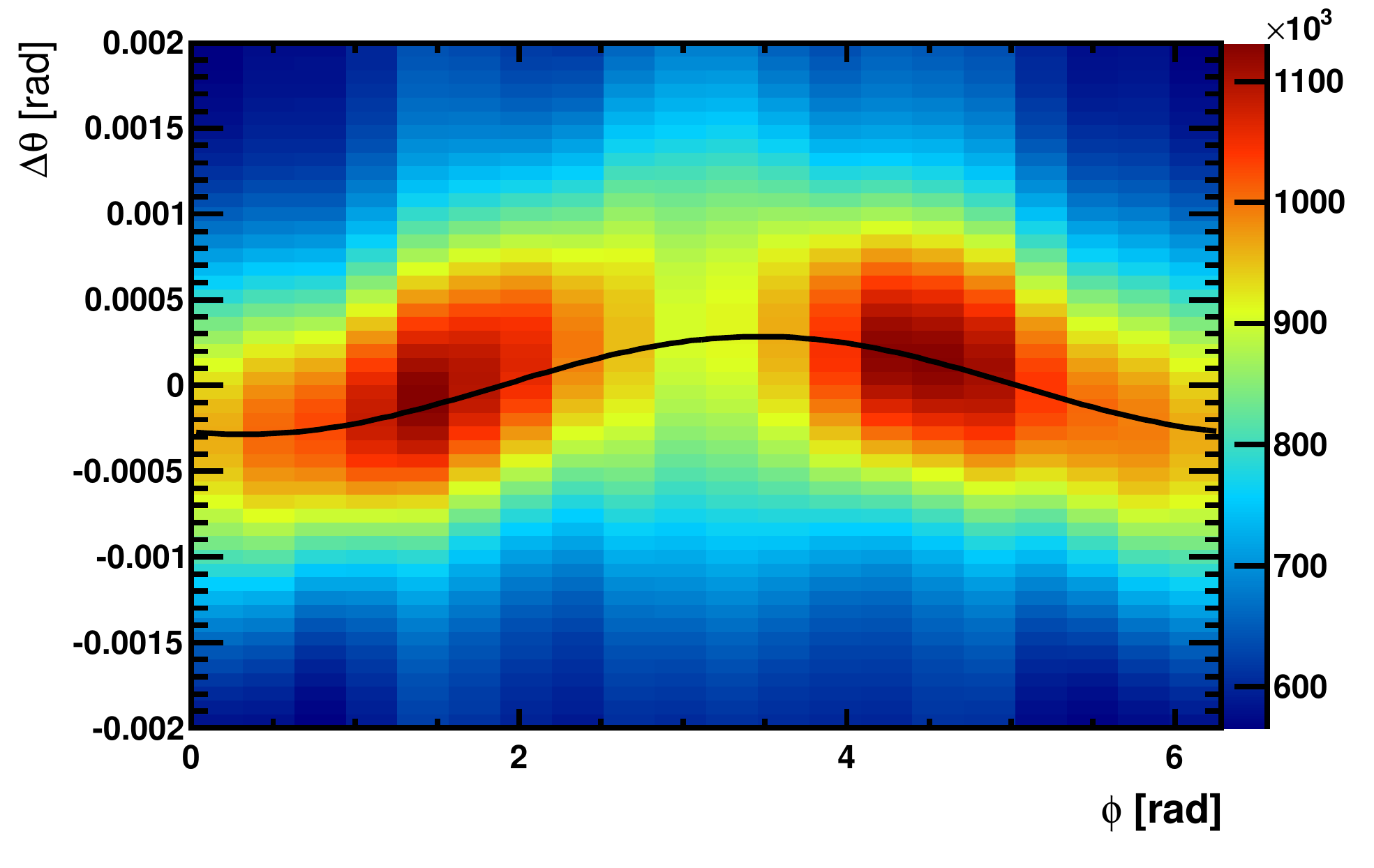}}
\\
\resizebox{0.45\textwidth}{!}{%
  \includegraphics{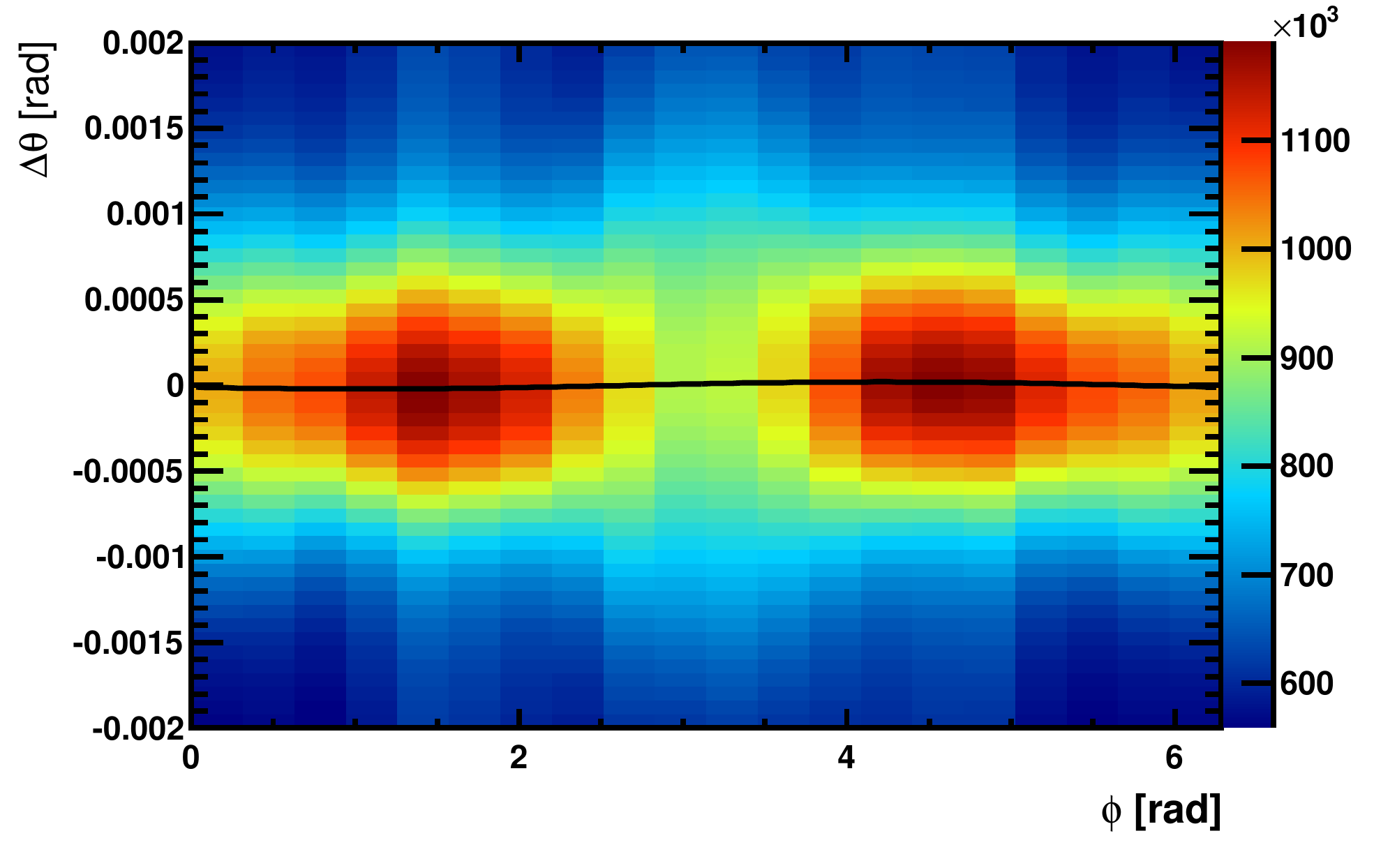}}
\caption{Difference between the measured and expected Cherenkov angle,
$\Delta\theta_C$ plotted as a function of the azimuthal angle $\phi$
and fitted with $\theta_{x} \cos(\phi)
+ \theta_{y}\sin(\phi)$, for one side of the RICH\,2 detector~\cite{LHCb-DP-2012-003}.
The upper plot is prior to alignment, and 
shows a dependency of the angle $\theta_{C}$ 
on the angle $\phi$. The bottom plot is after the alignment correction,
 and $\Delta \theta_{C}$ is uniform in $\phi$.}
\label{fig:RichMirrorAlign}
{\vskip-0.4cm}\end{figure}

The Hybrid Photon Detector is used to detect Cherenkov
photons. As shown in Fig.~\ref{fig:HPD},  
the photoelectron produced at the photocathode
is accelerated by a high voltage of up to 20~kV onto
a reverse-biased pixellated silicon detector, with a 
de-magnification factor of about 5~\cite{LHCb-DP-2012-003}.
The HPD anode images are affected by 
the magnetic and electric fields,
and have been observed to move and change their size,
possibly due to changes in these residual fields when the high voltage 
is cycled each LHC fill. 
Such changes could degrade the reconstruction of the Cherenkov angle and
affect the PID performance. 
Therefore the centre and radius of all the HPD images 
are calibrated run-by-run. 
Figure~\ref{fig:RichHPDCal} shows the calibration process. 
First, the centre of the image is cleaned to eliminate ion feedback.
Then a Sobel filter 
is used to detect the edges of the image that are fitted to 
determine the centre and the radius of the image, which are 
used by the RICH reconstruction in the final stage of the software
trigger. 
As only the raw HPD data needs to be decoded, more than 500~Hz
of events are processed run-by-run. 

\subsection{Alignment of the RICH mirror system}
\begin{figure}[!tb]
\centering
\resizebox{0.45\textwidth}{!}{%
\includegraphics{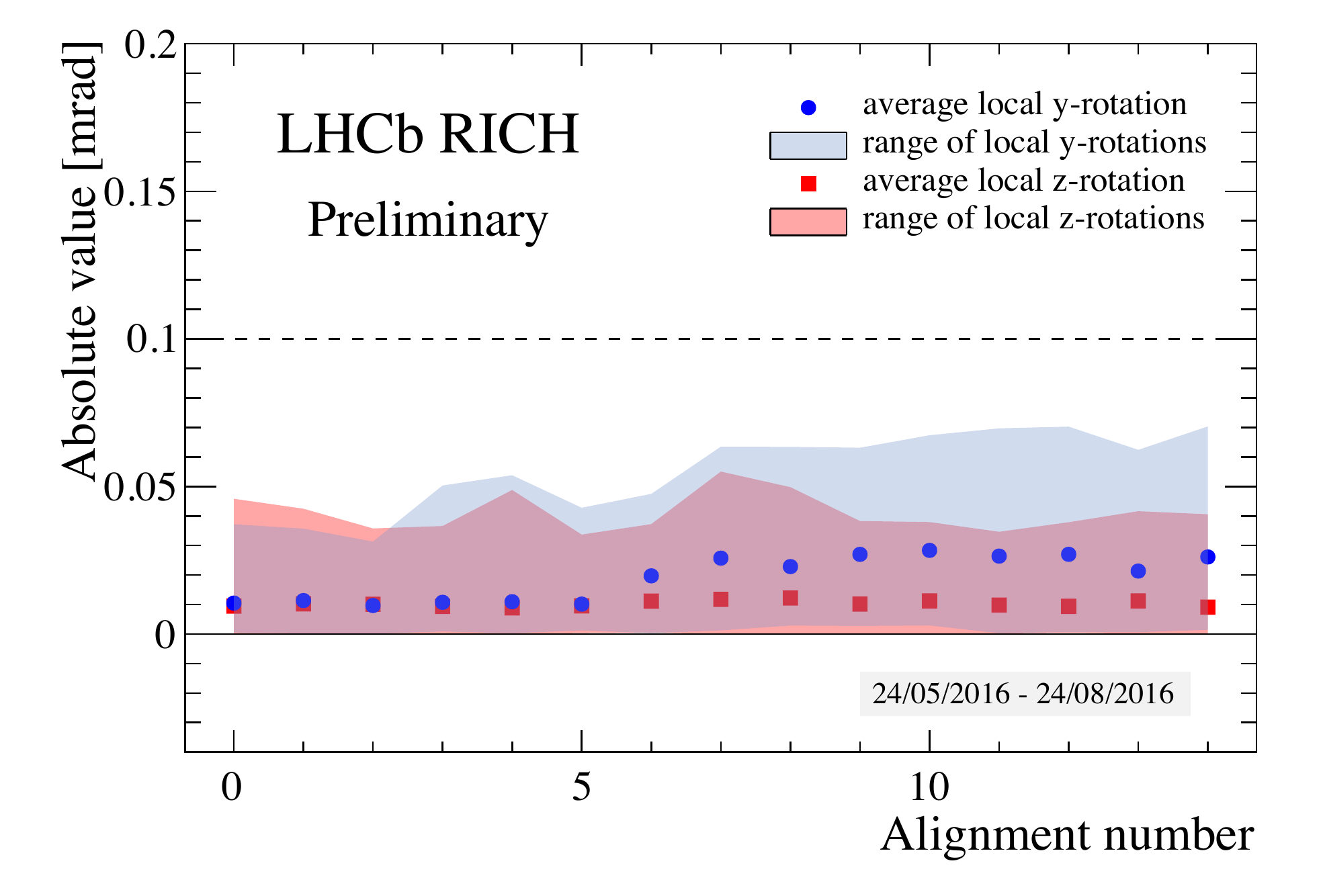}}
\\
\resizebox{0.45\textwidth}{!}{%
\includegraphics{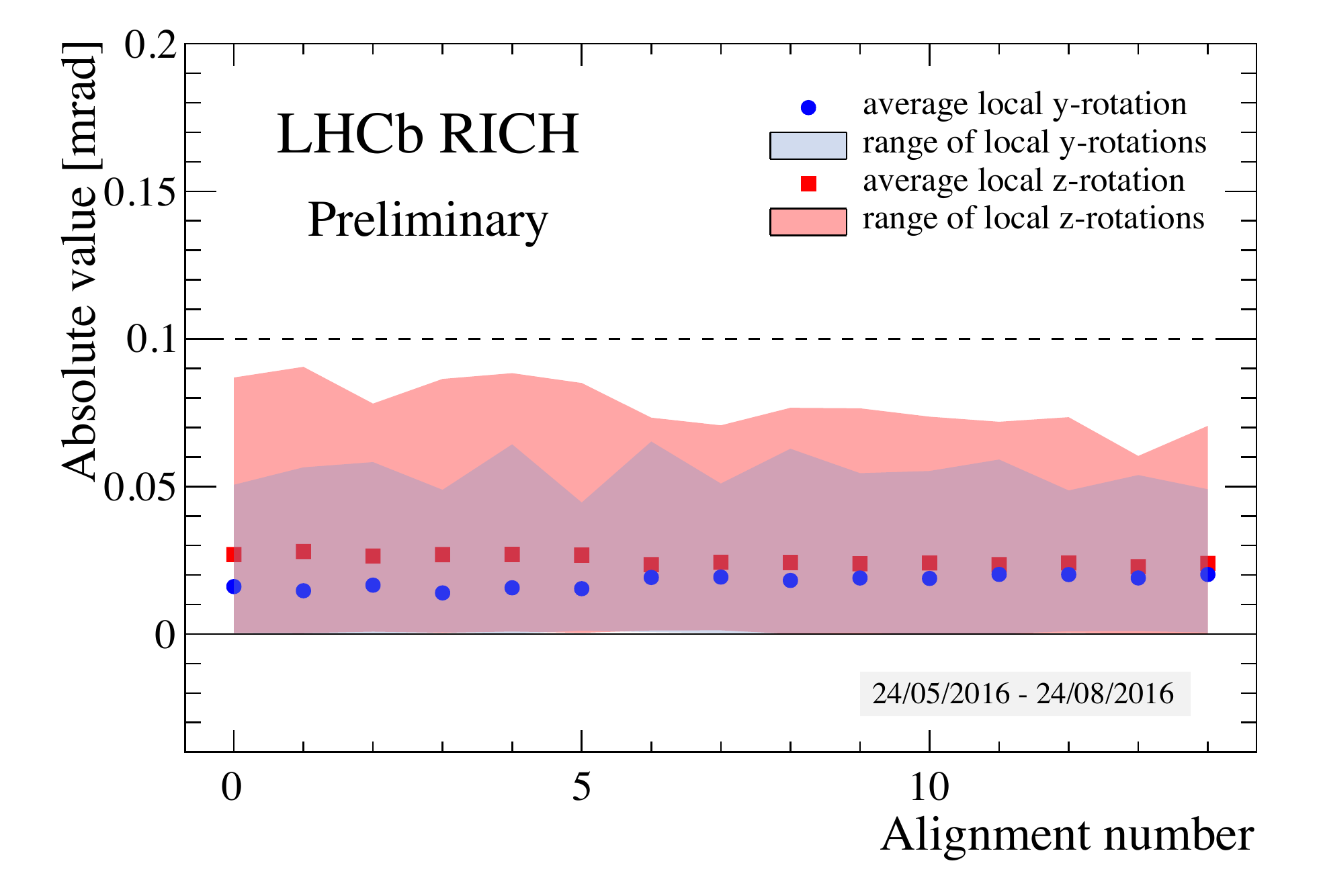}}
\caption{Time dependence of the mirror alignment parameters for the
  RICH detector downstream the magnet for the 2016 data sample, 
  (upper) spherical mirror, (bottom) secondary mirror. The shadow regions
  show the range of alignment parameters for all the mirror pairs,
  the points show the average value for all the mirror pairs.  
}
\label{fig:RichMirrorPerf}
{\vskip-0.8cm}\end{figure}

The Cherenkov photons emitted by the charged particles passing through 
the RICH detectors are focused onto the photon-detector
plane by the spherical and secondary mirrors. In case of misalignment
the centre of Cherenkov ring would not correspond to the intersection 
point of the charged track, and this would introduce a dependence 
of the difference between the measured and expected Cherenkov angle 
on the azimuthal angle of the ring, as shown in
Fig.~\ref{fig:RichMirrorAlign}. 
The alignment constants for each
mirror are determined by the fit of the Cherenkov angle difference
as a function of the azimuthal angle on the ring. 
The correlation between the different mirror pairs is also taken into
account. 
The procedure is evaluated by an iterative procedure implemented in
a dedicated framework, which makes it possible to run the alignment
in parallel using about 1800 nodes of the software trigger farm.  
The alignment of the RICH mirror system  
has been found to be stable enough to not affect the
PID performance, as shown in Fig.~\ref{fig:RichMirrorPerf},
and it runs routinely as a monitoring task.

\subsection{Time alignment}
\begin{figure}[!tb]
\centering
\resizebox{0.4\textwidth}{!}{%
\includegraphics{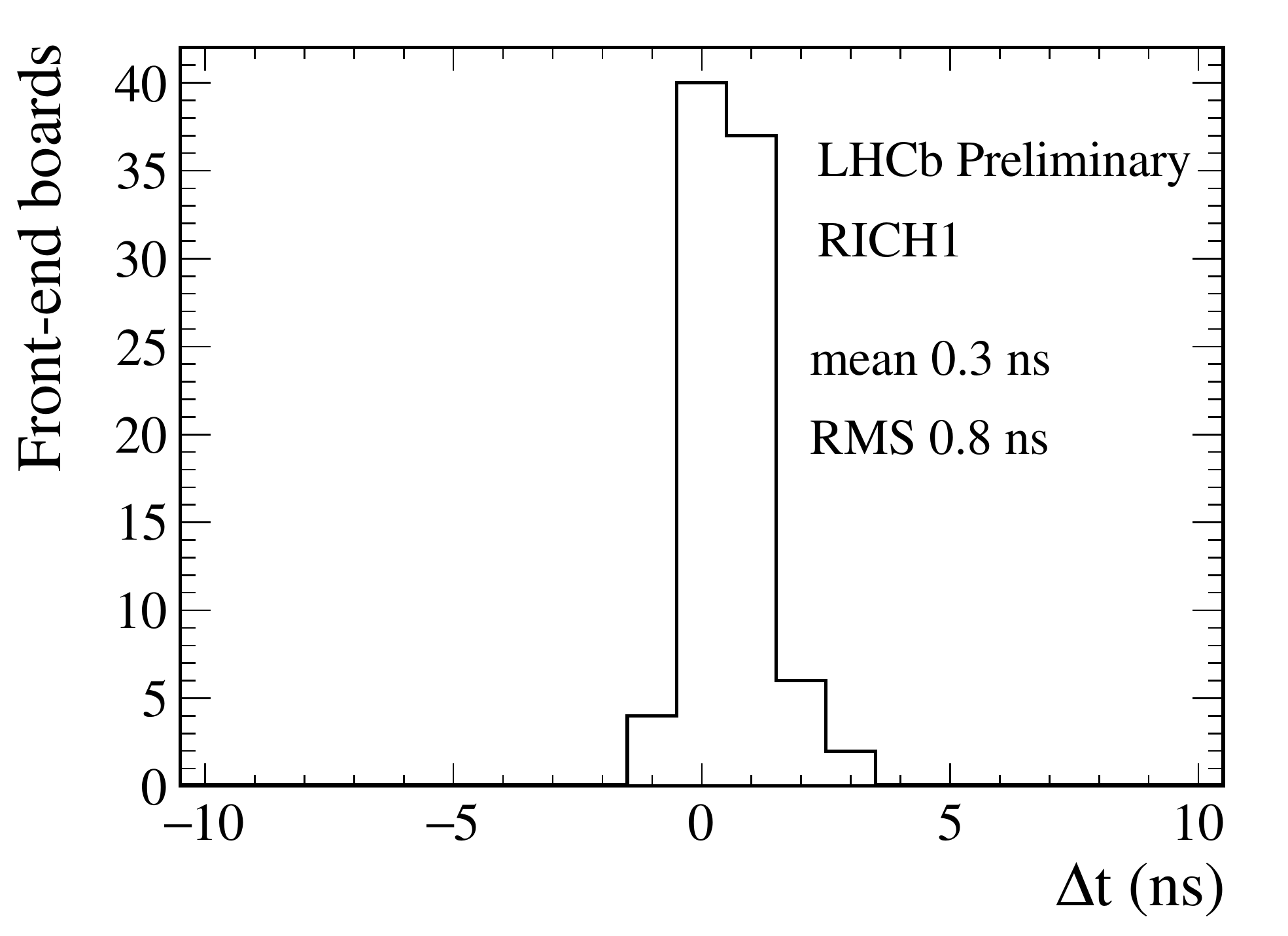}}
\\
\resizebox{0.4\textwidth}{!}{%
\includegraphics{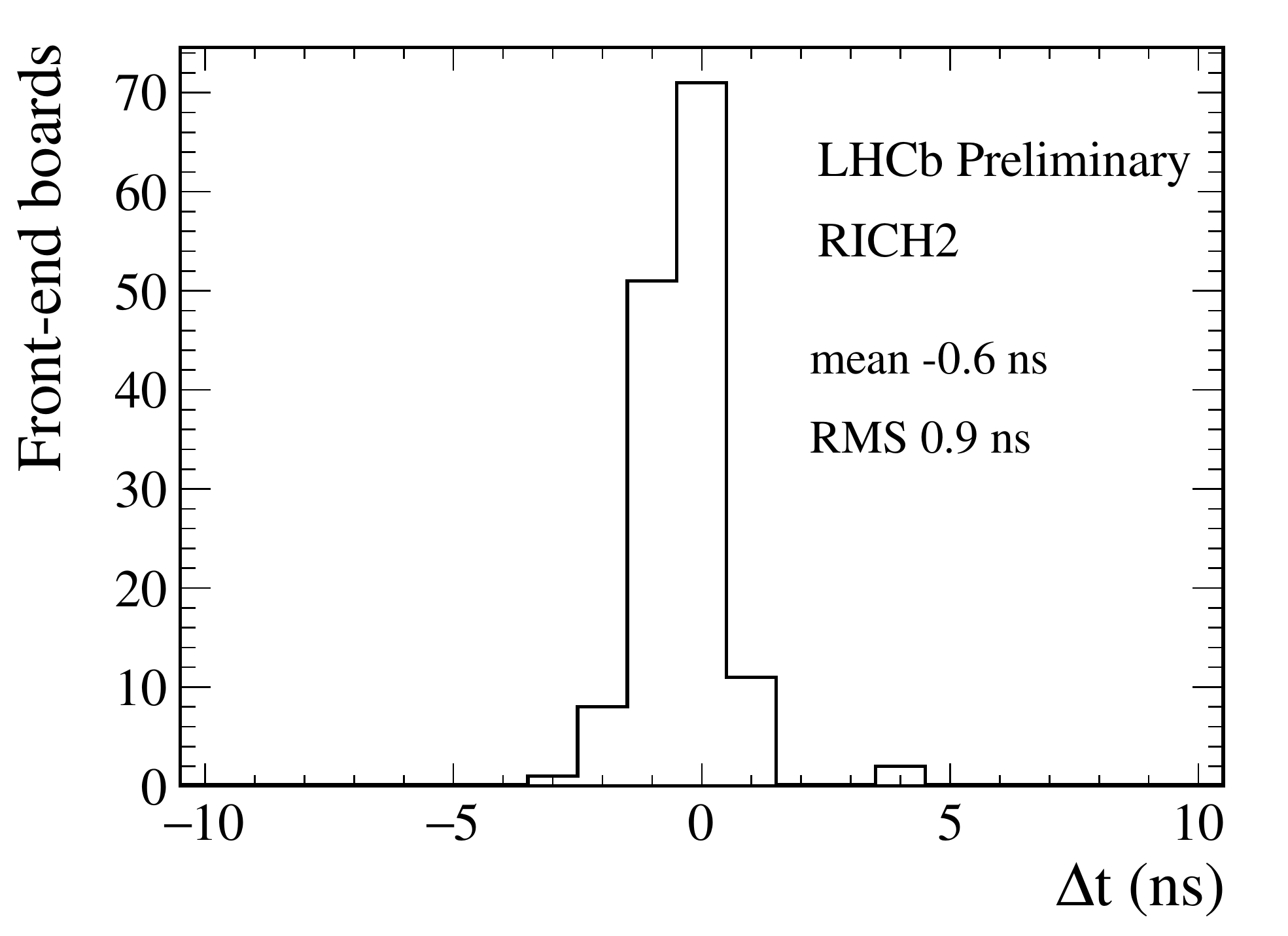}}
\caption{Distribution of the midpoints of timing scans in the upstream
  RICH (upper) and downstream RICH
  (bottom) after time alignment with pp collisions.}
\label{fig:RichTimAlign}
{\vskip-0.2cm}\end{figure}

In order to maximise the photon collection efficiency of the 
LHCb RICH detectors, the
HPD readout must be synchronised with the LHC bunch crossing to within
a few nanoseconds. 
The initial time alignment was performed in the absence of beam using
a pulsed laser, and has been improved further with dedicated timing
scan data taken during physics collisions. 
As shown in Fig.~\ref{fig:RichTimAlign}, all the HPDs have been time
aligned to about 1 ns.

\section{Performance}
\subsection{Stability of the RICH performance}

\begin{figure}[!ht]
\centering
\resizebox{0.45\textwidth}{!}{%
\includegraphics{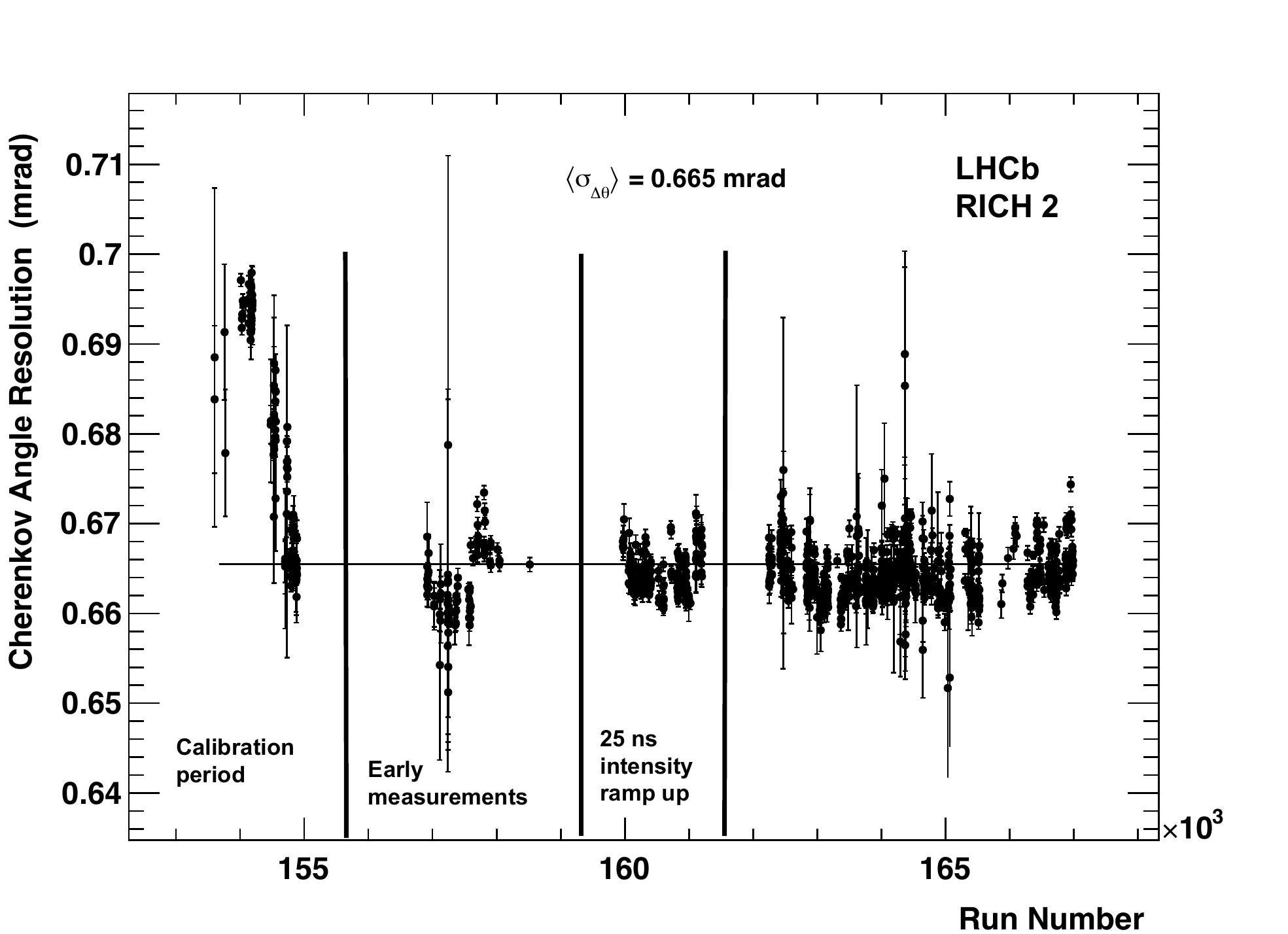}}
\caption{Time dependence of the Cherenkov angle resolution for the
  RICH detector downstream the magnet for the 2015 data sample.}
\label{fig:RichPerf}
{\vskip-0.2cm}\end{figure}

The stability of the RICH performance is monitored by the Cherenkov
angle resolution obtained with the same RICH mirror alignment and with
the dedicated run-by-run calibrations. It has been found to be stable
for the full 2015 data-taking, as shown in Fig.~\ref{fig:RichPerf}.

\subsection{Particle identification performance}
\begin{figure}[ht]
\centering
\resizebox{0.45\textwidth}{!}{%
\includegraphics{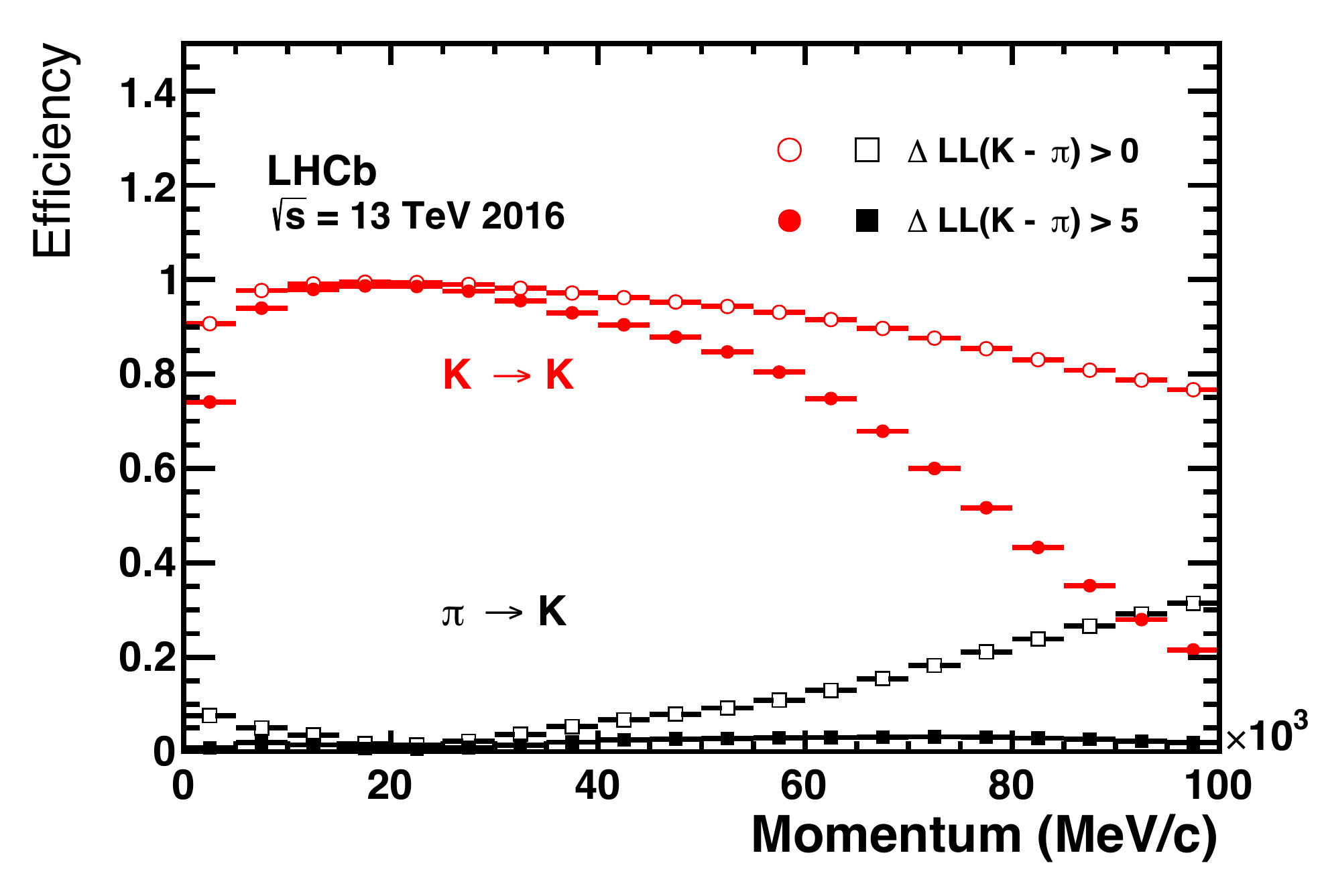}}
\caption{Kaon identification efficiency and pion misidentification
  rate measured on data as a function of track momentum. Two different
  $\Delta\mbox{log}\mathcal{L}(K-\pi)$ requirements have been imposed on the samples,
  resulting in the open and filled marker distributions, respectively.}
\label{fig:RichPIDPerf}
{\vskip-0.2cm}\end{figure}

Figure~\ref{fig:RichPIDPerf} shows 
the kaon efficiency (kaons identified as kaons)
and pion misidentification (pions misidentified as kaons), as a
function of particle momentum, obtained from imposing two different
requirements on this distribution in the Run-II data. 
One can see that the RICH detectors provide excellent particle 
identification. 
Further details on the particle identification performance in the
Run-II data-taking of the LHCb experiment can be found in
Ref.~\cite{Rich2016_Antonis}. 

\section{Summary}
Novel real-time calibration and alignment of the LHCb RICH detectors
have been implemented in the Run-II data-taking of the LHCb
experiment. This includes the run-by-run calibration of the refractive
index of the RICH radiators, and that of the Hybrid Photon Detector
image, and the alignment of the RICH mirror system. 
The real-time calibration and alignment works well, and provides
excellent particle identification for the online trigger and offline analysis.

\end{document}